\documentclass[12pt]{article}
\begin{document}
\title{A Model for Neutrinos}
\author{B.G. Sidharth\\
G.P. Birla Observatory \& Astronomical Research Centre\\
B.M. Birla Science Centre, Adarsh Nagar, Hyderabad - 500 063
(India)}
\date{}
\maketitle
\begin{abstract}
Neutrinos are the weirdest of elementary particles, nearly massless
and chiral. They oscillate between flavours, furthermore. We try to
find a rationale for this maverick behaviour.
\end{abstract}
\section{Introduction}
In 1930 Wolfgang Pauli regretted introducing the concept of the
neutrino on the grounds that it would not be observable. However,
thereafter as is well known, it was indeed observed in the context
of weak interactions, even though it has continued to remain
nature's most enigmatic particle. What is known is that its mass
could be $\sim 10^{-8}$ the electron mass, even though what has
actually been observed todate is the mass differences between
different flavours of neutrinos. These particles are fermions which
travel with nearly the speed of light. There are some $10^{90}$ of
them in the universe and more recently a huge cosmic background of
neutrinos has been observed \cite{weiler}. Unlike usual fermions,
the neutrino is described by the two component Weyl equation and
thus exhibits chirality, one of its weird characteristics. John
Wheeler has stressed that unless General Relativity can describe a
neutrino, in particular, its unification with Quantum Theory would
not be possible \cite{mwt}. Very recently the Gran Sasso Lab in
Italy has reported that what started up at LHC, Geneva as a beam of
$\mu$ neutrinos had switched flavours to appear as $\tau$ neutrinos
at Gran Sasso. Could such effects answer the problem of deficit of
solar neutrinos, is a question that comes up.
\section{Two Dimensionality}

We now observe that the neutrinos are in some sense two dimensional
entities and this would explain their maverick features. Firstly,
the Dirac equation in two space dimensions can represent the
neutrino as can be easily verified. Indeed this was pointed out by
the author nearly twenty years back \cite{bgsjsp,bgscamcstr} and
more recently this again came to be of interest because of the work
on graphene, which loosely is a two dimensional sheet of graphite.
Here we come across the phenomena of
quasi particles with distinctly neutrino like properties \cite{x,y,z}.\\ \\

Specifically, we first linearize the relativistic energy momentum
relation,
\begin{equation}
E^2 = p^2 c^2 + m^2_0 c\label{e1}
\end{equation}
and write down the corresponding quantum mechanical equation as
\begin{equation}
H \psi = [c \vec{\alpha} \cdot \vec{p} + \beta m_0 c^2]
\psi\label{e2}
\end{equation}
where
$$\vec{p} \equiv \frac{\hbar}{\imath} \left(\frac{\partial}{\partial
x} \frac{\partial}{\partial y}\right) \, \mbox{and} \, H \equiv
\frac{\hbar}{\imath} \frac{\partial}{\partial t}$$

Multiplying (\ref{e2}) by $H$ on the left side and $(c \vec{\alpha}
\cdot \vec{p} + \beta m_0 c^2)$ on the right side and comparing with
equation (\ref{e1}) we get,
\begin{equation}
(\alpha^\imath \alpha^j + \alpha^j \alpha^\imath ) = 2
\delta^{\imath j} , \alpha^\imath \beta + \beta \alpha^\imath = 0,
\beta^2 = 1 , \imath j = 1,2\label{e3}
\end{equation}
Equation (\ref{e3}) is satisfied if the set $(\vec{\alpha}, \beta)$
is the set of Pauli matrices.\\
Thus equation (\ref{e2}) represents a two component Fermion.
Relativistic covariance of this equation can be easily verified. The
neutrino is such a two component Fermion \cite{schweber}. It is also
well known that the reason the neutrino which satisfies the Dirac
equation, or its special case, the Weyl equation has only two
components is that its rest mass $m_0$ vanishes.\\
So if a Fermion can somehow be confined to two dimensions, for
example to the surface of a thin strip of negligible thickness, for
example graphene, then it should behave effectively like the
massless and parity-violating
neutrino.\\ \\

We must also remember that the cosmic background neutrinos are at
the temperature $T \approx 3^\circ K$ that is this collection is
ultra cold and as such nearly mono energetic. We can argue that this
leads to a bosonic behaviour as follows. Our starting point is the
well known formula for the occupation number of a Fermion
gas\cite{huang}
\begin{equation}
\bar n_p = \frac{1}{z^{-1}e^{bE_p}+1}\label{8je9}
\end{equation}
where, $z' \equiv \frac{\lambda^3}{v} \equiv \mu z \approx z$
because, here, as can be easily shown $\mu \approx 1,$
$$v = \frac{V}{N}, \lambda = \sqrt{\frac{2\pi \hbar^2}{m/b}}$$
\begin{equation}
b \equiv \left(\frac{1}{KT}\right), \quad \mbox{and} \quad \sum \bar
n_p = N\label{8je10}
\end{equation}
Let us consider in particular a collection of Fermions which is
somehow made nearly mono-energetic, that is, given by the
distribution,
\begin{equation}
n'_p = \delta (p - p_0)\bar n_p\label{8je11}
\end{equation}
where $\bar n_p$ is given by (\ref{8je9}).\\
This is not possible in general - here we consider a special
situation of a collection of nearly mono-energetic particles in
equilibrium which is the case with the Cosmic Background neutrinos.\\
By the usual formulation we have,
\begin{equation}
N = \frac{V}{\hbar^3} \int d\vec p n'_p = \frac{V}{\hbar^3} \int
\delta (p - p_0) 4\pi p^2\bar n_p dp = \frac{4\pi V}{\hbar^3} p^2_0
\frac{1}{z^{-1}e^{\theta}+1}\label{8je12}
\end{equation}
where $\theta \equiv bE_{p_0}$.\\
It must be noted that in (\ref{8je12}) there is a loss of dimension
in momentum space, due to the $\delta$ function in (\ref{8je11}) -
in fact such a fractal two dimensional situation would in the
relativistic case lead us to anomalous behaviour described elsewhere
\cite{bgscosmic}. One way of looking at this is that dimensionality
itself is connected to the virial distribution of velocities and in
this special
situation, there is hardly such a velocity spread.\\ \\

Indeed this two dimensionality aspect has entered physics through
the holographic principle in the most recent Quantum Gravity
approaches \cite{witten}. Here to put it simply three dimensionality
is perceptional as in the three dimensional image of a hologram. All
the detail is contained in the two dimensional image. In the case of
Black Holes, all information content is confined to its surface.
Indeed the universe itself
can be characterized as a Black Hole \cite{tduniv}.\\ \\

Furthermore in 1996 it was argued by the author (and A.D. Popova)
that the universe is asymptotically two dimensional. A quick way to
see this would be by observing the dynamics of the flat rotation
curves of galaxies \cite{bgspopova}. Essentially, observations
reveal, that for galaxies at large distances $R$, their mass goes as
$$M \propto R^n, \quad \mbox{where} \, n \approx 2$$
indicating two dimensionality.\\ \\

All this suggests that the neutrino can be treated in the above
context as a two dimensional object obeying the two dimensional
Dirac or Weyl equation.\\ \\

There is yet another way of looking at all this. It is well known,
in the theory of the Dirac equation, that if we construct wave
packets of positive energy solutions alone (or negative energy
solutions alone) \cite{bd}, we get
\begin{equation}
\psi^{(+)} (x,t) = \int \frac{d^3 p}{(2 \pi \hbar)^{1/2}}
\sqrt{\frac{mc^2}{E}} \sum_{\pm s} b (p,s) u(p,s) e^{-\imath
p_{\mu}x\mu/\hbar}\label{3.23}
\end{equation}
The expectation of the velocity operator is given by
\begin{equation}
\bf{J}^{(+)} = \int \psi^{(+)\dag} con\psi^{(+)} d^3 x\label{3.25}
\end{equation}
which leads to
\begin{equation}
\bf{J}^{(+)} = \langle c \alpha\rangle_+ = \left\langle \frac{c^2
p}{E}\right\rangle_+ = \langle V_{gp}\rangle_+\label{3.29}
\end{equation}
where $\langle  \rangle_+$ denotes expectation value with respect to
a positive-energy packet. Clearly (\ref{3.29}) shows that the
velocity is luminal. Indeed Dirac himself noticed this as soon as he
had derived the relativistic electron equation \cite{diracpqm} and
rationalized that all measurements are averages over a short
interval of time of the order of the Compton time, in which the
Zitterbewegung effects are smoothened out and we get a sub luminal
value for the velocity. To put it another way, a wave packet
constructed of solutions of one sign of energy alone move with
luminal velocities. We need a wave packet consisting of both
positive and negative energies to get the usual sub luminal
velocities and mass. This has been commented upon \cite{bgsultra}.\\
Turning to the neutrino we can now describe the neutrino as being
made up of one sign of energy solutions alone, leading to its
luminal speeds and near vanishing mass. This also means that Dirac
spinor equation which consists of four components, two representing
positive solutions and two representing negative solutions becomes a
two component, that is positive only or negative only spinor. This
of course is the well known Weyl equation which represents neutrinos
as two component objects. All this explains several features
including the near masslessness and  the chirality.\\
In any case it can be argued that a particle localized in space
would be described by a packet that contains both positive energy
and negative energy solutions. Those particles described by only
positive or only negative energy solutions would be luminal and
therefore massless. Interestingly this would also mean that the
Compton wavelength of such particles, in this case neutrinos would
be infinite (or in practise very large) \cite{bd,fb,greiner,bgsultra}.\\
Interestingly the cosmic neutrino background has recently been used
by the author to explain the cosmological constant \cite{bgsnap}.
The starting point is the cold Fermi degenerate gas for which we
have
\begin{equation}
p^3_F  = \hbar^3 (N/V)\label{ex}
\end{equation}
where we can deduce this by using the fact that the ground state of
such a Fermi assembly is such that the neutrinos occupy the lowest
possible energy levels, while all the levels up to the Fermi energy
$\epsilon$ or $e_F$ are filled up. That is these neutrinos fill the
sphere of radius $p_F$. So we have
\begin{equation}
\frac{V}{h^3} \int_{e_p < e_F} d^3 p = N\label{XX}
\end{equation}
Remembering that we have,
$$e_F  \sim p^2_F / m$$
(\ref{ex}) follows.\\
Feeding in the known neutrino parameter, viz., \cite{ruffini} $N
\sim 10^{90}$ we get from the above, the consistent neutrino mass
$\sim 10^{-3} eV$ \cite{tegmark} and the background temperature $T
\sim 1^\circ K$ as $KT$ is nothing but the Fermi energy $e_F$. More
recently there has been hope that neutrinos can also exhibit the
ripples of the early Big Bang and in fact,
Trotta and Melchiorri claim to have done so \cite{trotta}.\\
It may be mentioned that there is growing evidence for the cosmic
background neutrinos \cite{weiler}. The GZK photo pion
process seems to be the contributing factor.\\
Next using the expression for the Fermi energy
\begin{equation}
\mbox{Fermi \, Energy}\, = \frac{N^{5/3} \hbar^2}{m_\nu R^2} = M
\Lambda R^2\label{XA}
\end{equation}
where $M$ is the mass of the universe, $R$ its radius $\sim
10^{27}cm,$ and $\Lambda$ is the cosmological constant.\\
We now get from (\ref{XA}),
\begin{equation}
\Lambda \sim 10^{-37} sec^{-2}\label{XB}
\end{equation}
(\ref{XB}) describes the correct order of the cosmological
constant.\\
It may be noted that by conventional arguments we get a wrong value
of the cosmological constant that is $10^{120}$ orders of magnitude
of the actually observed cosmological constant.

\end{document}